\documentclass[article,twocolumn, superscriptaddress]{revtex4}
\usepackage{graphicx}
\usepackage{epstopdf}
\usepackage{dcolumn}
\usepackage{bm}
\usepackage{pinlabel}
\usepackage{amsmath}
\usepackage{xcolor}
\usepackage{comment}
\usepackage{lipsum}

\begin{document}
\title{Multiple magneto-ionic regimes in Ta/Co$_{20}$Fe$_{60}$B$_{20}$/HfO$_{2}$}

\author{R. Pachat}
\affiliation{Centre de Nanosciences et de Nanotechnologies, CNRS, Universit\'e Paris-Saclay, 91120 Palaiseau, France}

\author{D. Ourdani}
\affiliation{Laboratoire des Sciences des Proc\'ed\'es et des Mat\'eriaux, CNRS-UPR 3407, Universit\'e Paris 13, Sorbonne Paris Cit\'e, 93430 Villetaneuse, France}

\author{J. W. van der Jagt}
\affiliation{Spin-Ion technologies, C2N, 10 Boulevard Thomas Gobert, 91120 Palaiseau, France}

\author{M.-A. Syskaki}
\affiliation{Singulus Technologies AG, Hanauer Landstrasse 103, 63796 Kahl am Main, Germany}

\author{A. Di Pietro}
\affiliation{Istituto Nazionale di Ricerca Metrologica, Strada delle Cacce 91, 10135 Torino, Italy}

\author{Y. Roussign\'e}
\affiliation{Laboratoire des Sciences des Proc\'ed\'es et des Mat\'eriaux, CNRS-UPR 3407, Universit\'e Paris 13, Sorbonne Paris Cit\'e, 93430 Villetaneuse, France}

\author{S. Ono}
\affiliation{Central Research Institute of Electric Power Industry, Yokosuka,\mbox{Kanagawa 240-0196,} Japan}

\author{M. S. Gabor}
\affiliation{Center for Superconductivity, Spintronics and Surface Science, Physics and Chemistry Department, Technical University of Cluj-Napoca, Cluj-Napoca RO-400114, Romania}

\author{M. Ch\'erif}
\affiliation{Laboratoire des Sciences des Proc\'ed\'es et des Mat\'eriaux, CNRS-UPR 3407, Universit\'e Paris 13, Sorbonne Paris Cit\'e, 93430 Villetaneuse, France}

\author{G. Durin}
\affiliation{Istituto Nazionale di Ricerca Metrologica, Strada delle Cacce 91, 10135 Torino, Italy}

\author{J. Langer}
\affiliation{Singulus Technologies AG, Hanauer Landstrasse 103, 63796 Kahl am Main, Germany}

\author{M. Belmeguenai}
\affiliation{Laboratoire des Sciences des Proc\'ed\'es et des Mat\'eriaux, CNRS-UPR 3407, Universit\'e Paris 13, Sorbonne Paris Cit\'e, 93430 Villetaneuse, France}

\author{D. Ravelosona}
\affiliation{Centre de Nanosciences et de Nanotechnologies, CNRS, Universit\'e Paris-Saclay, 91120 Palaiseau, France}
\affiliation{Spin-Ion technologies, C2N, 10 Boulevard Thomas Gobert, 91120 Palaiseau, France}

\author{L. Herrera Diez}
\email{liza.herrera-diez@c2n.upsaclay.fr}
\affiliation{Centre de Nanosciences et de Nanotechnologies, CNRS, Universit\'e Paris-Saclay, 91120 Palaiseau, France}


\begin{abstract}
	In Ta/CoFeB/HfO$_{2}$ stacks a gate voltage drives, in a nonvolatile way, the system from an underoxidized state exhibiting in-plane anisotropy (IPA) to an optimum oxidation level resulting in perpendicular anisotropy (PMA) and further into an overoxidized state with IPA. The IPA$\,\to\,$PMA regime is found to be significantly faster than the PMA$\,\to\,$IPA regime, while only the latter shows full reversibility under the same gate voltages. The effective damping parameter also shows a marked dependence with gate voltage in the IPA$\,\to\,$PMA regime, going from 0.029 to 0.012, and only a modest increase to 0.014 in the PMA$\,\to\,$IPA regime. The existence of two magneto-ionic regimes has been linked to a difference in the chemical environment of the anchoring points of oxygen species added to underoxidized or overoxidized layers. Our results show that multiple magneto-ionic regimes can exist in a single device and that their characterization is of great importance for the design of high performance spintronics devices. 
	
\end{abstract}
\maketitle
Controlling magnetic properties with electric (E) fields is of great importance for spintronics applications due to its potential for lowering power consumption in novel memory prototypes. It has been shown that E-field induced charge accumulation can lead to important changes in magnetic anisotropy \cite{weisheit_electric_2007,daalderop_magnetocrystalline_1991,nakamura_giant_2009,duan_surface_2008}, which can be used to reliably control domain wall (DW) pinning and velocities \cite{bernand-mantel_electric-field_2013, liu_electric_2017}, and assist the magnetization switching in magnetic tunnel junctions \cite{wang_electric-field-assisted_2012} in ferromagnetic (FM) metallic films. More complex mechanisms like switching between ferromagnetic and skyrmionic states \cite{hsu_electric-field-driven_2017, schott_skyrmion_2017} have been also demonstrated, where the E-field control of the Dzyaloshinskii-Moriya Interaction (DMI) \cite{srivastava_large-voltage_2018, koyama_electric_2018} is also at play.\\
In addition to the effects of charge accumulation, magneto-ionics can also provide nonvolatility and a more extended effect over the entire magnetic layer, unlike charge accumulation, limited by electrostatic screening \cite{zhang_spin-dependent_1999}. Large magneto-ionic effects in magnetic anisotropy can be achieved, including a spin reorientation transition, which applied to DW motion can be used to create nonvolatile and very efficient domain wall traps \cite{bauer_voltage-controlled_2013, bauer_magneto-ionic_2015}. In addition, magneto-ionic control of spin accumulation \cite{mishra_electric-field_2019} and DMI \cite{herrera_diez_nonvolatile_2019} have been demonstrated as well as DW chirality switching in a controlled oxygen atmosphere \cite{chen_large_2020}, which is of capital importance for implementing E-field assisted spin-orbit torques in devices. In these ionic materials, the migration of oxygen \cite{bauer_magneto-ionic_2015,gilbert_structural_2016} or hydrogen species \cite{tan_magneto-ionic_2019} is at the heart of the ionic effects observed, which have recently been shown to offer operation speeds close to 1 ms \cite{lee_fast_2020}.\\
The interest in nonvolatility for spintronics applications goes hand in hand with magneto-ionic reversibility. Recently, studies have focused on the magneto-ionic reversibility variations between
\begin{figure}
	\vspace{0 cm}
	\includegraphics[width=7cm]{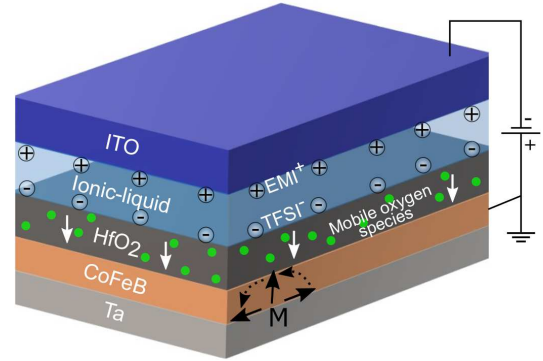}
	\caption{Graphic representation of the magneto-ionic stack covered with the ionic liquid [EMI]$^+$[TFSI]$^-$ gate. The gate voltage induces the motion of oxygen species in HfO$_{2}$.}
	\label{device}
\end{figure}
oxides \cite{fassatoui_physical_2020}, pointing out the importance of the differences in the mechanisms governing ionic conduction. A deep understanding of the mechanisms of reversibility in magneto-ionics is therefore needed to design high performing magneto-electric devices. In this study, we unveil a higher degree of complexity in HfO$_{2}$ based devices, where two distinct magneto-ionic regimes have been identified in a single structure. In addition, we also shine light on the magneto-ionic effects on the effective damping parameter $\alpha_{\mathrm{eff}}$, largely overlooked in the literature and of great importance for fast magnetization dynamics. We show that $\alpha_{\mathrm{eff}}$ can be significantly reduced by inducing ionic motion and that its lowest value (0.012) coincides with the appearance of PMA.\\
The magnetic materials used in this study are amorphous Ta(5\,nm)/Co$_{20}$Fe$_{60}$B$_{20}$(1\,nm)/HfO$_{2}$(3\,nm) films grown by magnetron sputtering, all samples investigated here were cut from the same wafer. The ionic liquid (IL) [EMI][TFSI] (1-Ethyl-3-methylimidazolium bis(trifluoromethanesulfonyl)imide) is added to the surface of the film to incorporate the IL gate, which can provide high E-fields that have been shown to induce ionic motion in a variety of materials \cite{jeong_oxygen-vacancy_2013, shi_resistance_2013, leighton_ionic-control_2019}. The thickness of the IL is macroscopic, in the range of several hundreds of $\mu$m. However, the effective IL thickness is estimated taking into account only the distance over which an electric double layer is formed at the side of each one of the electrodes, which is 1 nm \cite{ono_ionic-liquid_2008}. A counter electrode, a glass substrate coated with a 100 nm thick indium tin oxide (ITO) layer is subsequently placed on top of the IL. The size of the E-field biased area is of about 0.25 cm$^{2}$. Samples were stored in air and at room temperature before conducting the gating experiments if not indicated otherwise. The hysteresis loops have been measured by anomalous Hall effect using a bias current of 400 $\mu$A. All magnetic states presented here are nonvolatile, all measurements were conducted after switching off the gate voltage. A graphic representation of the device geometry is presented in Fig. \ref{device}.\\
In the initial state, before exposure to the gate voltage, the
\begin{figure}
	\vspace{0 cm}
	\includegraphics[width=8cm]{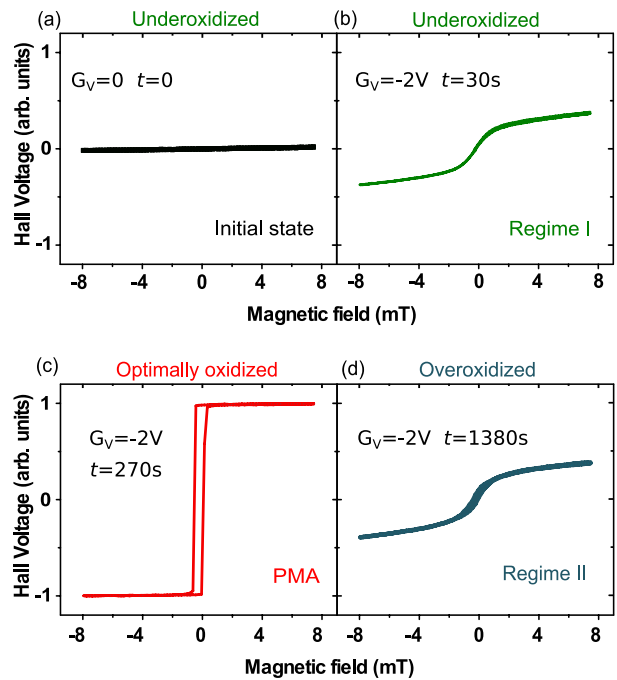}
	\caption{A progressive oxidation is induced upon exposure to a gate voltage G$_{V}$=-2 V. Different exposure times \textit{t} drive the system from IPA (a, initial state) through regime I (b) into PMA (c) and back to IPA through regime II (d).}
	\label{oxidation}
\end{figure}
magnetic layers are underoxidized, exhibiting in-plane magnetic anisotropy (IPA). Under the action of a negative gate voltage applied to the top ITO electrode a first regime (I) is identified, in which IPA transitions into PMA, followed by a second regime (II) where further oxidation drives PMA into IPA. This entire anisotropy evolution is presented in Fig.\ref{oxidation} with the corresponding cumulative biasing times for each case, these values correspond to the accumulated time of all previous operations. The gate voltage G$_{V}$=-2 V was applied for a total, accumulated time of \textit{t}=1380 s. It is well known that thin magnetic films can exhibit a window of oxidation levels at the interface with an oxide which promotes PMA, while under/over oxidation will result in IPA \cite{manchon_analysis_2008}. Oxygen species are also known to migrate from the HfO$_{2}$ toward the CoFeB layer under gate voltages, as reported for other  HfO$_{2}$ based devices \cite{herrera_diez_nonvolatile_2019, nagata_oxygen_2010}, therefore oxygen migration is thought to be at the heart of the effects observed here. In the following, we will describe in detail the two magneto-ionic regimes and the impact of ionic diffusion on the effective damping parameter $\alpha_{\mathrm{eff}}$. 

\section{Magneto-ionic regimes I and II}
\begin{figure}
	\vspace{0 cm}
	\includegraphics[width=8.5cm]{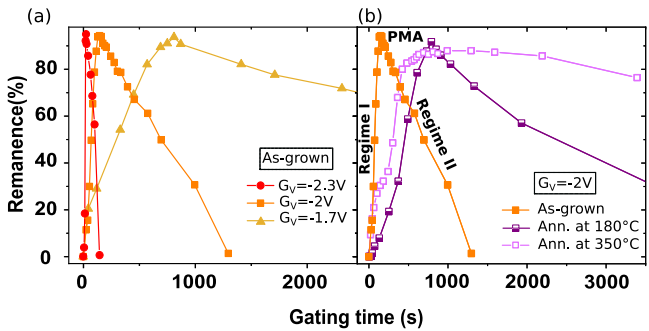}
	\caption{Remanence as a function of gating time in (a) as-grown samples under G$_{V}$=-2.3 V (circles), G$_{V}$=-2 V (squares), and G$_{V}$=-1.7 V (triangles). (b) Samples annealed at 180 °C (half open squares) and 350 °C (open squares) under G$_{V}$=-2 V.}
	\label{time}
\end{figure}

\begin{figure}
	\vspace{0 cm}
	\includegraphics[width=8cm]{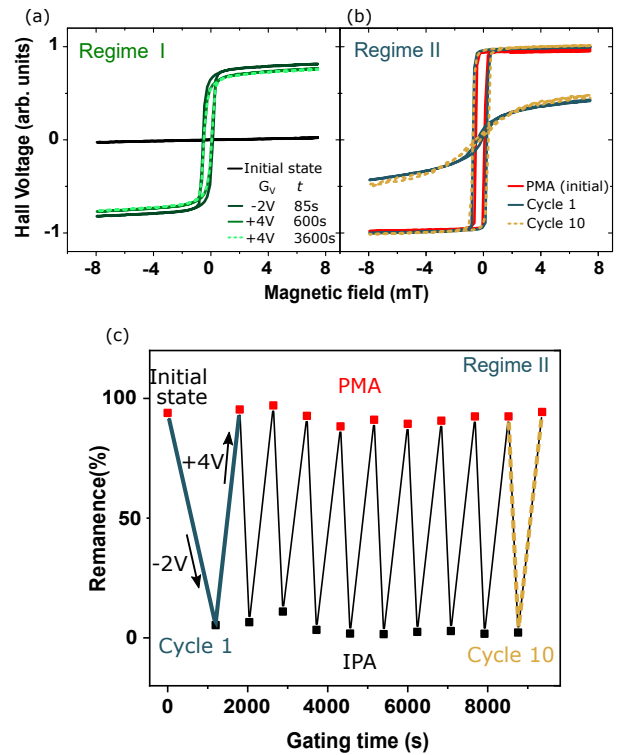}
	\caption{(a) Suppressed reversibility of the IPA$\,\to\,$PMA transition under G$_{V}$=+4 V in regime I. (b) Hysteresis loops for the reversible PMA$\,\to\,$IPA cycles 1 (solid line) and 10 (dotted line) in regime II. (c) Remanence as a function of the reversibility cycle number in regime II.}
	\label{reversibility}
\end{figure}

Fig.\ref{time} (a) shows the time dependence of the out-of-plane remanence percentage, where 100\% corresponds to the Hall voltage at zero applied magnetic field in the PMA state, for the entire anisotropy rang going from the initial underoxidized IPA state through regime I (positive slope), PMA (the point of highest remanence), and regime II (negative slope) to finish in an overoxidized IPA state. As-grown samples have been investigated under three different constant gate voltages of G$_{V}$=-2.3 V (circles), G$_{V}$=-2 V (squares), and G$_{V}$=-1.7 V (triangles). The speed of the entire process critically increases for higher gate voltages, responding to a stronger driving force for ionic motion. For all applied gate voltages, the magneto-ionic process in regime I is significantly faster than in regime II. 
This difference in the time evolution as a function of gate voltage between regime I and II could be linked to a difference in the chemical environment, and thus of anchoring, of the oxygen species interacting with the CoFeB layer in regimes I and II. In regime I, the energy barrier associated with adding oxygen species to the underoxidized CoFeB interface may be lower than in regime II, where oxygen species have to be added to an already optimally oxidized CoFeB interface.\\
Measurements were also conducted using underoxidized samples that were annealed before applying a gate voltage of -2 V, these measurements are presented in Fig. \ref{time} (b). An underoxidized sample was annealed at 180 $^{\circ}$C (open squares) for one hour to reduce moisture without inducing the crystallization of the CoFeB. Under G$_{V}$=-2 V, the speed of the ionic process in both regimes I and II was critically reduced, while still showing the same difference previously discussed between regimes I and II. This has been linked to a contribution to the mobile oxygen species coming from air humidity, a feature observed in ionic systems based on hydrogen mobility \cite{tan_magneto-ionic_2019}. A sample annealed at 350 $^{\circ}$C (half open squares) for one hour was also investigated under G$_{V}$=-2 V. These annealing conditions are known to induce the formation of crystallized grains in CoFeB/MgO stacks, where the MgO layer acts as a crystallization template \cite{yuasa_characterization_2005,wang_-situ_2009}, and the migration of B towards the Ta substrate \cite{sinha_borondiffusion_2015}. In CoFeB/HfO$_{2}$, a similar behaviour could be tentatively proposed, where B diffusion \cite{bart_hfo2_2017} and the formation of crystalline grains would have an impact in ionic mobility. Fig. \ref{time} shows that annealing has an impact in ionic dynamics in this system, regime I is slower than in the as-grown sample while regime II is significantly slower with respect to all other samples investigated. As mentioned earlier, the chemical environment of the mobile oxygen species at their anchoring points in the CoFeB layer is likely to be non equivalent in regimes I and II. An already slower regime II further reducing its velocity in crystallized samples may be an indication that ions in this regime do not only occupy binding sites at the already oxygen rich interface but that they could also penetrate into the layer. In this context, the crystalline structure of the CoFeB grains could hinder the diffusion of ions inside the layer with respect to the amorphous case, significantly reducing the speed and efficiency of the incorporation of oxygen species. However, a detailed study of the impact of annealing on the crystalline structure of Ta/CoFeB/HfO$_{2}$ is needed to fully understand the changes observed in the magneto-ionic behaviour.\\
Regimes I and II do not only show two distinct speeds of the magneto-ionic process but they also show different reversibility behaviors. The reversibility of the effects of a gate voltage of -2 V has been tested with positive gate voltages going up to +4 V in regimes I and II of as-grown samples. Higher voltages have not been applied due to the electrochemical limitations of the ionic liquid \cite{seidemann_iontronics_2017}. The effects of the application of +4 V for 10 minutes and up to one hour to a magnetic state close to PMA in regime I are shown in Fig.\ref{reversibility} (a), where only minor changes are induced in the magnetic state. The same effect is found for other intermediate states in regime I and also for the fully perpendicular state, positive gate voltages up to +4 V and long exposures can not induce a recovery of the initial under oxidized state. Regime II shows an entirely different behaviour; a fully reversible transition between IPA and PMA can be observed. Fig. \ref{reversibility}(b) shows the Hall voltage hysteresis loops corresponding to reversibility cycles number 1 and 10, while Fig. \ref{reversibility}(c) shows the remanence variation of the entire series. The ten cycles show a switching of the remanence between nearly 0 $\%$ (IPA) and 100 $\%$ (PMA) under gate voltages of -2 V and +4V, respectively. It is interesting to notice that in the first cycle, the PMA$\,\to\,$IPA transition is induced by the application of -2 V for 1200 s, while for all subsequent cycles a shorter time of 240 s is required for this transition to occur. For the IPA$\,\to\,$PMA transition an exposure time of 600 s has been found to be sufficient for all cycles. This shows that the magneto-ionic process most likely undergoes
a first 'activation' phase in which the first diffusion of ions involves an additional energy barrier. This could be related to the energy cost of a first detachment of the ions from their original anchoring points as well as to the formation of ionic conduction channels in the CoFeB layer, after which the ions can be moved reversibly by using significantly lower exposure times.\\
Regime I is therefore showing a relatively faster dynamics than in regime II under G$_{V}$=-2 V and a highly suppressed reversibility at G$_{V}$=+4 V, while in regime II, a slower dynamics and full reversibility are observed. This may be linked to a high degree of stability of the final position of the ions within regime I, which would create a high energy barrier for the inverse process, suppressing reversibility under a gate voltage of +4 V. In regime II, although the energy barrier remains asymmetric between the PMA$\,\to\,$IPA and PMA$\leftarrow$IPA transitions, a gate voltage of +4 V is sufficient to revert the effects of a gate voltage of -2 V. This could be linked to a weaker anchoring of the mobile oxygen species in regime II. 

\section{Effective damping parameter}
\begin{figure}
	\vspace{0 cm}
	\includegraphics[width=8cm]{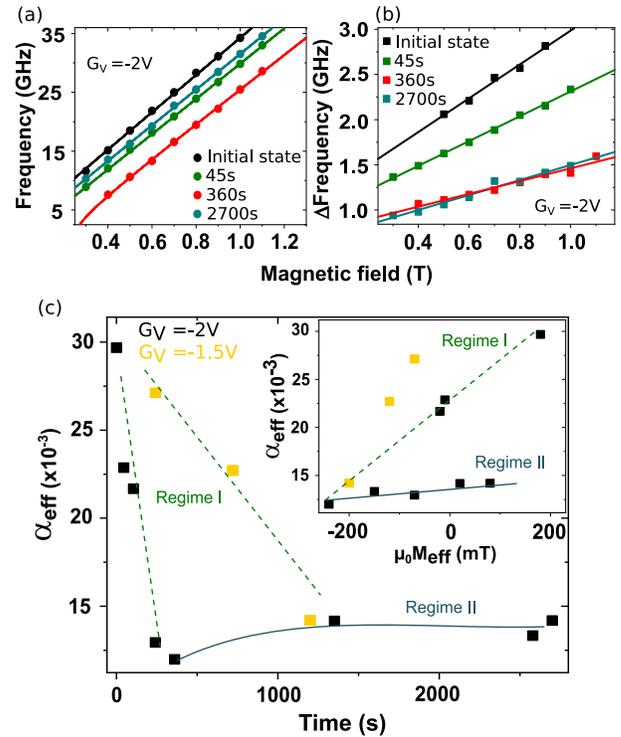}
	\caption{(a) BLS average frequency and (b)  linewidth as a function of magnetic field for an as-grown (black circles) sample, and for samples exposed to G$_{V}$=-2 V for 45 s (green circles), 350 s (red circles) and 2700 s (cyan circles), and the corresponding fitting lines. (c) $\alpha_{\mathrm{eff}}$ dependence on exposure time to G$_{V}$=-2 V (black squares) and G$_{V}$=-1.5 V (yellow squares) and on $\mu _{0} M_{\mathrm{eff}}$ (inset). Lines are a guide to the eye.}
	\label{damping}
\end{figure}

In this section, we will discuss the impact of magneto-ionics on the effective damping parameter $\alpha_{\mathrm{eff}}$. A series of samples, exposed to gate voltages -2 V and -1.5 V, has been investigated by Brillouin light scattering (BLS) in the Damon-Eshbach geometry after removal of the IL gate. The in-plane magnetic field dependence of the average frequency of the Stokes and anti-Stokes spin-wave frequencies was measured at a wavenumber  \textit{k}=8.08 $\mu$m$^{-1}$.  Fig. \ref{damping}(a) shows this dependence for the as-grown sample (black circles) and for samples exposed to G$_{V}$=-2 V for 45 s (underoxidized, green circles), 360 s (PMA, red circles) and 2700 s (overoxidized, cyan circles). A theoretical modelling of this dependence has been performed according to \cite{belmeguenai_interfacial_2015} resulting in the fitting lines shown in Fig. \ref{damping}(a). From this fitting, the values of the effective magnetisation $M_{\mathrm{eff}} = M_{\mathrm{s}}-H_{k}$ have been obtained, where $M_{\mathrm{s}}$ and $H_{k}$ are the saturation magnetisation and the anisotropy field, respectively. Fig. \ref{damping}(b) shows the corresponding plots of the BLS frequency full linewidth ($\Delta F$) as a function of the magnetic field applied in the plane of the sample and the corresponding fitting lines to: $\Delta F = 2\alpha_{\mathrm{eff}} (\gamma/2\pi )H + \Delta F_{0} $, where $\alpha_{\mathrm{eff}}$ can be extracted from the slope. The dependence of $\alpha_{\mathrm{eff}}$ on exposure time to a gate voltage of -2 V (black squares) and -1.5 V (yellow squares) is presented in Fig. \ref{damping}(c), where the inset shows 
also the dependence on $\mu _{0} M_{\mathrm{eff}}$

. In regime I, the value of $\alpha_{\mathrm{eff}}$ rapidly decreases from 0.029 for the as-grown sample to 0.012 for the PMA state. As shown earlier, the change from G$_{V}$=-2 V to G$_{V}$=-1.5 V critically decreases the speed of the process. In regime II, a relatively small increase in $\alpha_{\mathrm{eff}}$ to 0.014 is observed, however, it does not show to have a marked dependence on the exposure time. This difference between regimes I and II is also expressed in the $\alpha_{\mathrm{eff}}$ vs. $\mu _{0} M_{\mathrm{eff}}$ dependence, where a monotonic decrease is seen in regime I, in contrast with the much less pronounced increase seen for regime II in a similar $\mu _{0} M_{\mathrm{eff}}$ range.\\
It is important to mention that the values of $\alpha_{\mathrm{eff}}$ obtained from the slopes of the data points in Fig. \ref{damping}(b) contain not only the intrinsic contribution from the Gilbert damping parameter but also an extrinsic contribution. This extrinsic contribution can be associated with a variety of sources including spin pumping \cite{nakayama_geometry_2012, belmeguenai_ferromagnetic-resonance-induced_2018, belmeguenai_exchange_2017}, due to the proximity with a heavy metal with high spin-orbit coupling, two-magnon scattering \cite{kuanr_extrinsic_2005} or the existence of sample inhomogeneities such as a distribution of anisotropy values across the sample \cite{capua_determination_2015,diez_enhancement_2019}. The extrinsic contribution can add a relatively small factor to the value of the Gilbert damping or in more extreme cases it can dominate the $\Delta F$ vs $H$ dependence resulting in the loss of linearity \cite{capua_determination_2015, diez_enhancement_2019}.\\
The damping parameter in CoFeB has been shown to depend on the oxidation level at the interface. In CoFeB/Gd/MgO stacks, the thin Gd layer serving as an oxygen sink has been shown to modulate the level of oxidation of an overoxidized CoFeB. In this system, $\alpha_{\mathrm{eff}}$ shows a non-monotonic dependence on the thickness of the Gd layer where a minimum is reached for 0.6 nm \cite{yang_effect_2017}. However, this has been linked mostly to changes in the degree of homogeneity and quality of the interface. As mentioned, the proximity and quality of the interface with a high spin orbit coupling material has been shown to induce a spin-pumping contribution to the effective damping which shows as an inverse proportionality between $\alpha_{\mathrm{eff}}$ and the thickness of the magnetic film \cite{ikeda_perpendicular-anisotropy_2010, benguettat-el_mokhtari_interface_2020}. This additional damping contribution has been quantified for Ta by comparing symmetric MgO/CoFeB/MgO and Ta/CoFeB/Ta stacks to asymmetric Ta/CoFeB/MgO, where the Ta contribution dominates over the MgO contribution \cite{iihama_gilbert_2014}. This has also been shown in the case of Hf/CoFeB/MgO stacks, which show for a CoFeB thickness of 1.08 nm, $\alpha_{\mathrm{eff}}$ values that are higher than in MgO/CoFeB/MgO stacks by about a factor of five \cite{lourembam_thickness-dependent_2018}.\\
\begin{figure*}
	\vspace{0 cm}
	\includegraphics[width=17cm]{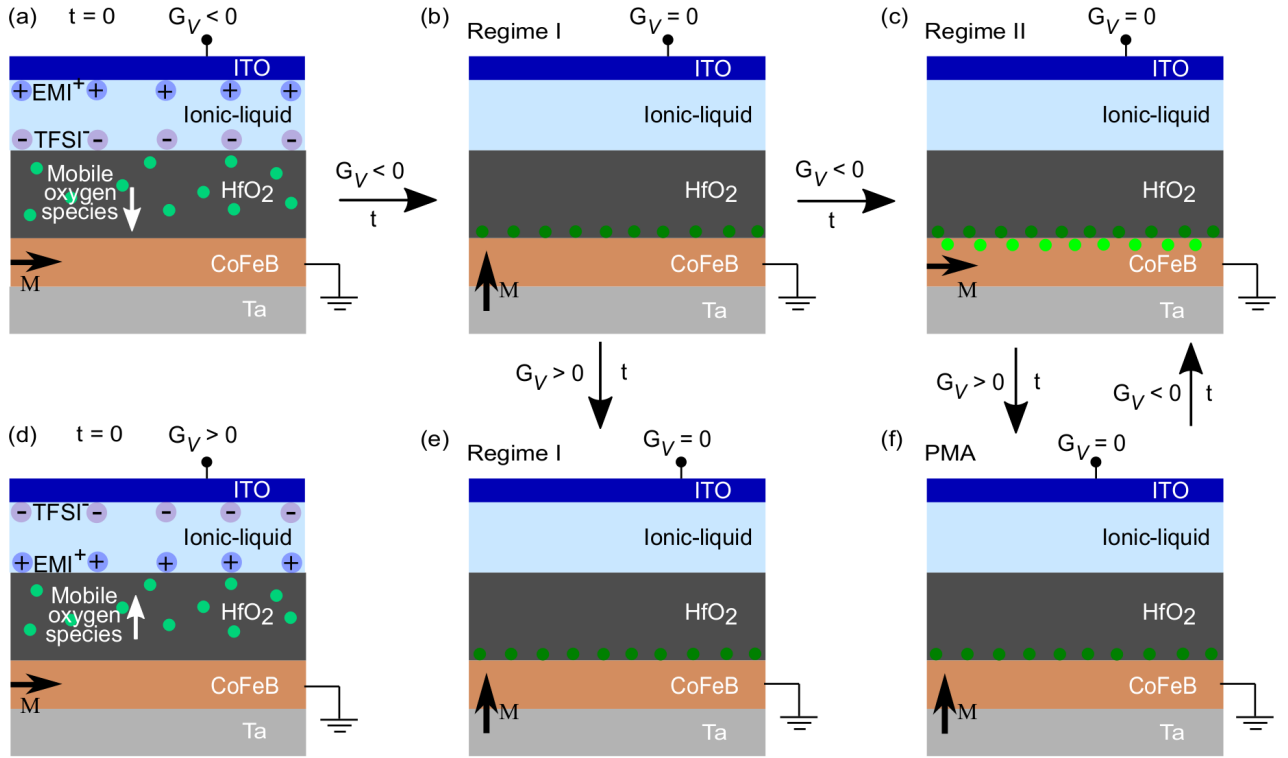}
	\caption{Proposed magneto-ionic mechanisms for regimes I and II. (a) For G$_{V}<$ 0 oxygen species migrate toward the CoFeB layer achieving an optimum surface coverage and PMA at the end of regime I (b). If G$_{V}<$ 0 continues to be applied, oxygen species will continue to be incorporated into the CoFeB layer in regime II (c) resulting in IPA. Under G$_{V}>$ 0 mobile oxygen species can migrate back to the top electrode (d). For the same applied G$_{V}$, there is no reversibility in regime I ((b) and (e)), while full reversibility is achieved in regime II ((c) and (f)).}
	\label{mechanism}
\end{figure*}
Taking into account these considerations, it can be concluded that in the present case the underoxidized samples could have a significantly larger contact surface with Hf in HfO$_{2}$ than in the optimally oxidised case. A progression in regime I toward PMA could therefore decrease the contact surface with Hf and significantly reduce the associated spin pumping contribution to $\alpha_{\mathrm{eff}}$. Interestingly, the changes in $\alpha_{\mathrm{eff}}$ are minor in regime II, which supports the idea mentioned earlier that in this regime oxygen species may migrate mostly into the layer rather than binding at the already optimally oxidized interface reached after regime I. This is considered as the main contribution to the observed variations in $\alpha_{\mathrm{eff}}$, however, a reordering of the interface structure throughout regime I is also a likely scenario. A sharper interface may be reached at the end of regime I, also leading to the highest PMA state, which could also play a role in the observed reduction of $\alpha_{\mathrm{eff}}$. 

\section{Discussion}
The proposed magneto-ionic mechanisms involved in regimes I and II are depicted in Fig. \ref{mechanism}.
For G$_{V}<$ 0, mobile oxygen species inside HfO$_{2}$, represented as green dots, migrate toward the CoFeB layer and after G$_{V}$ is switched off (b), they remain bound to the CoFeB, achieving an optimum surface coverage and completing the IPA$\,\to\,$PMA transition at the end of regime I. If G$_{V}<$ 0 continues to be applied, the magneto-ionic mechanism will move into regime II. After the voltage is switched off (c), the incorporated oxygen species will contribute to the PMA$\,\to\,$IPA transition. As the starting point of regime II is an optimally oxidized CoFeB surface, oxygen species incorporated in regime II will be less likely to bind at the same binding sites available in regime I and can potentially penetrate further into the layer (light green dots).\\
Under high enough positive gate voltages, mobile oxygen species are expected to migrate back to the top electrode (d). However, for the same applied G$_{V}>$ 0, reversibility is highly suppressed for oxygen species incorporated in regime I ((b) and (e)), while full reversibility is achieved in regime II ((c) and (f)). This discrepancy in the degree of reversibility of regimes I and II is attributed to the differences in the binding sites, and the associated anchoring strength, of the oxygen species incorporated in each regime. The difference in chemical environment for these two types of binding sites is proposed as the origin of magneto-ionic regimes I and II.\\
A preferential occupation of a limited number of binding sites at the CoFeB/HfO$_{2}$ interface in regime I could also explain the observed larger impact of ionic migration on the value of $\alpha_{\mathrm{eff}}$ with respect to regime II. As mentioned, the effects observed may be due to a decoupling of the CoFeB surface from the Hf atoms in HfO$_{2}$, mostly linked to ion incorporation at the surface of an underoxidized CoFeB rather than to ionic migration into the layer in regime II. This is valid in a scenario where ionic migration into the layer is limited to the vicinity of the CoFeB/HfO$_{2}$ interface and does not extend up to the Ta/CoFeB interface as seen in other systems where higher gate voltages are applied \cite{herrera_diez_nonvolatile_2019}. It could be anticipated that ionic migration up to the Ta/CoFeB interface could induce a significant decoupling from the Ta layer, an effect observed in Pt/Co reflected as a reduction of the DMI \cite{herrera_diez_nonvolatile_2019}, resulting in a further reduction of $\alpha_{\mathrm{eff}}$. Another important indication supporting the proposed mechanism is found in the measurements done on crystallized samples. In this case, the time evolution of magnetic anisotropy under a negative gate voltage is strikingly reduced in regime II compared to regime I. This could be due to an increased difficulty of ionic diffusion into a crystallized CoFeB layer in regime II, compared to ionic incorporation at the surface in regime I.\\
Surface and bulk components to the magneto-ionic effect have already been introduced in the literature, in particular, in Pd/Co/GdO$_{x}$ and Pt/Co/HfO$_{2}$ films, where ion migration deep down into the magnetic layer and beyond can cause partial or total irreversibility \cite{gilbert_structural_2016, herrera_diez_nonvolatile_2019}. As mentioned, in the present case the bulk component is thought to be limited only to the vicinity of the CoFeB/HfO$_{2}$ interface. The observed suppressed reversibility seen in regime I happens at low oxygen content, ruling out the anchoring of the oxygen species deep into the layer as a potential source of irreversibility. This is also well supported by the subsequent observation of regime II, exhibiting full reversibility, in the overoxidized state.\\
In this context, studies in the literature addressing the incorporation of oxygen species from the gas phase into crystalline Co and Fe systems provide an interesting perspective. In Co, the profile of the oxygen intake as a function of oxygen pressure has a distinct change in slope that is related to a two stage oxidation process. The first step involves the formation of CoO up to a defined surface coverage threshold, after which a higher oxygen and high temperature exposure leads to the formation of Co$_{3}$O$_{4}$ \cite{bridge_oxygen_1979}. In Fe crystals, a similar behaviour has been observed at room temperature. The oxygen sticking coefficient as a function of surface coverage is seen to decrease drastically up to a coverage of 0.5, where an inflection point occurs followed by a less pronounced decrease. The initial phase of rapid decrease has been linked to the filling of preferential empty binding sites, while additional oxygen incorporation leads to the inflection point identified as the onset of oxide growth. The oxygen incorporation is proposed to entail a progression from initially adsorbed oxygen species to the nucleation and expansion of a two dimensional layer of FeO at the sticking coefficient inflection point, where it remains fairly constant until the completion of the FeO layer formation. Further oxygen incorporation happens in the presence of FeO, which is responsible for the observation of a reduction in the sticking coefficient after the inflection point. This last process ultimately leads to the formation of Fe$_{2}$O$_{3}$/ Fe$_{3}$0$_{4}$ together with the appearance of a three dimensional oxide structure \cite{brundle_oxygen_1977, simmons_leed-aes_1975}.\\
In amorphous systems, the distinction between the well defined  oxide phases present in crystalline systems could be less evident, however, a distinction between purely surface incorporation and a subsequent oxygen intake more extended into the magnetic layer could be evidenced. The films used in this study are Fe rich, with a composition of Co$_{20}$Fe$_{60}$B$_{20}$, where variations of the Fe content at the interface with an oxide have been shown to drive large PMA variations \cite{herrera_diez_controlling_2015}. It is therefore worth considering a scenario where the formation of different Fe oxide types is at the origin of regimes I and II. Regime I is likely to reflect the initial oxygen adsorption and formation of FeO, which involves a change in valence from Fe$^{0}$ to Fe$^{2+}$. A significantly high energy barrier is therefore introduced for the inverse process which would largely suppress reversibility under a low energy stimulus. On the other hand, the full completion of a two dimensional FeO layer could well lead to an optimum and homogeneous oxidation at the interface, allowing for the highest PMA values observable in this system. As mentioned, further oxygen incorporation in the presence of FeO becomes less favourable, which is in line with the observed slower dynamics seen in regime II. The Fe$_{2}$O$_{3}$/ Fe$_{3}$O$_{4}$ oxide phase involves a mixture of Fe$^{2+}$ and Fe$^{3+}$ states, therefore it can be proposed that the path toward a Fe$^{2+}$ $\,\to\,$ Fe$^{3+}$ transition involves the progressive addition of oxygen species to the chemical environment of the Fe$^{2+}$ centres. This phase is thought to describe regime II, where a relatively low energy barrier is involved in the inverse process due to the conservation of the Fe$^{2+}$ valence state. An ultimate change to Fe$^{3+}$ would also be expected to introduce a large energy barrier for the inverse process and hinder reversibility, as seen in regime I. This is well in line with the observed behaviour under G$_{V}$= -3 V (120 s exposure time), where the system enters an irreversible IPA overoxidized state.

\section{Conclusion}                  
In conclusion, we have shown the existence of two distinct nonvolatile magneto-ionic regimes in Ta/CoFeB/HfO$_{2}$ stacks where oxygen species migrate under negative/positive gate voltages toward/away from the CoFeB layer. This voltage driven ionic motion induces first an IPA$\,\to\,$PMA transition in regime I, corresponding to a transition from an underoxidized to an optimally oxidized state. In regime II, a PMA$\,\to\,$IPA transition occurs and it is correlated to a transition from an optimally oxidized to an overoxidized state. Regime I shows a much faster dynamics and highly suppressed reversibility for positive gate voltages with respect to regime II. In addition, it also shows a marked decrease of the effective damping parameter from 0.029 to 0.012 compared to the relatively small increase to 0.014 observed in regime II. The existence of regimes I and II is proposed to be the result of a difference in the binding strength of the migrated oxygen species, that can be correlated with different binding sites on the surface and inside the CoFeB layer, respectively.\\
The results presented here reveal the complexity of magneto-ionics and the importance of a deep understanding of the ionic mechanisms involved in order to design robust and reliable devices for spintronics applications. This system could be easily transferred to a solid state device where the two regimes could be probed at higher gate voltages in order to re-evaluate reversibility and explore fast operation times. This will allow to choose an IPA to PMA transition, either in regime I or II, that best fits the requirements for practical applications and to design strategies to favour operation in one of the two magneto-ionic regimes. 

\begin{acknowledgments}
	\noindent We gratefully acknowledge financial support from the European Union H2020 Program (MSCA ITN grant No. 860060), from the French National Research Agency (project ELECSPIN), and the program PHC Sakura. This work was in part supported by JSPS KAKENHI Grant Number 20H05304.
	
\end{acknowledgments}


\begin{thebibliography}{42}
\expandafter\ifx\csname natexlab\endcsname\relax\def\natexlab#1{#1}\fi
\expandafter\ifx\csname bibnamefont\endcsname\relax
\def\bibnamefont#1{#1}\fi
\expandafter\ifx\csname bibfnamefont\endcsname\relax
\def\bibfnamefont#1{#1}\fi
\expandafter\ifx\csname citenamefont\endcsname\relax
\def\citenamefont#1{#1}\fi
\expandafter\ifx\csname url\endcsname\relax
\def\url#1{\texttt{#1}}\fi
\expandafter\ifx\csname urlprefix\endcsname\relax\def\urlprefix{URL }\fi
\providecommand{\bibinfo}[2]{#2}
\providecommand{\eprint}[2][]{\url{#2}}

\bibitem[{\citenamefont{Weisheit et~al.}(2007)\citenamefont{Weisheit, Fahler,
		Marty, Souche, Poinsignon, and Givord}}]{weisheit_electric_2007}
\bibinfo{author}{\bibfnamefont{M.}~\bibnamefont{Weisheit}},
\bibinfo{author}{\bibfnamefont{S.}~\bibnamefont{Fahler}},
\bibinfo{author}{\bibfnamefont{A.}~\bibnamefont{Marty}},
\bibinfo{author}{\bibfnamefont{Y.}~\bibnamefont{Souche}},
\bibinfo{author}{\bibfnamefont{C.}~\bibnamefont{Poinsignon}},
\bibnamefont{and} \bibinfo{author}{\bibfnamefont{D.}~\bibnamefont{Givord}},
\bibinfo{journal}{Science} \textbf{\bibinfo{volume}{315}},
\bibinfo{pages}{349} (\bibinfo{year}{2007}), ISSN \bibinfo{issn}{0036-8075,
	1095-9203},
\urlprefix\url{https://www.sciencemag.org/lookup/doi/10.1126/science.1136629}.

\bibitem[{\citenamefont{Daalderop et~al.}(1991)\citenamefont{Daalderop, Kelly,
		and Schuurmans}}]{daalderop_magnetocrystalline_1991}
\bibinfo{author}{\bibfnamefont{G.~H.~O.} \bibnamefont{Daalderop}},
\bibinfo{author}{\bibfnamefont{P.~J.} \bibnamefont{Kelly}}, \bibnamefont{and}
\bibinfo{author}{\bibfnamefont{M.~F.~H.} \bibnamefont{Schuurmans}},
\bibinfo{journal}{Physical Review B} \textbf{\bibinfo{volume}{44}},
\bibinfo{pages}{12054} (\bibinfo{year}{1991}), ISSN \bibinfo{issn}{0163-1829,
	1095-3795},
\urlprefix\url{https://link.aps.org/doi/10.1103/PhysRevB.44.12054}.

\bibitem[{\citenamefont{Nakamura et~al.}(2009)\citenamefont{Nakamura,
		Shimabukuro, Fujiwara, Akiyama, Ito, and Freeman}}]{nakamura_giant_2009}
\bibinfo{author}{\bibfnamefont{K.}~\bibnamefont{Nakamura}},
\bibinfo{author}{\bibfnamefont{R.}~\bibnamefont{Shimabukuro}},
\bibinfo{author}{\bibfnamefont{Y.}~\bibnamefont{Fujiwara}},
\bibinfo{author}{\bibfnamefont{T.}~\bibnamefont{Akiyama}},
\bibinfo{author}{\bibfnamefont{T.}~\bibnamefont{Ito}}, \bibnamefont{and}
\bibinfo{author}{\bibfnamefont{A.~J.} \bibnamefont{Freeman}},
\bibinfo{journal}{Physical Review Letters} \textbf{\bibinfo{volume}{102}},
\bibinfo{pages}{187201} (\bibinfo{year}{2009}), ISSN
\bibinfo{issn}{0031-9007, 1079-7114},
\urlprefix\url{https://link.aps.org/doi/10.1103/PhysRevLett.102.187201}.

\bibitem[{\citenamefont{Duan et~al.}(2008)\citenamefont{Duan, Velev,
		Sabirianov, Zhu, Chu, Jaswal, and Tsymbal}}]{duan_surface_2008}
\bibinfo{author}{\bibfnamefont{C.-G.} \bibnamefont{Duan}},
\bibinfo{author}{\bibfnamefont{J.~P.} \bibnamefont{Velev}},
\bibinfo{author}{\bibfnamefont{R.~F.} \bibnamefont{Sabirianov}},
\bibinfo{author}{\bibfnamefont{Z.}~\bibnamefont{Zhu}},
\bibinfo{author}{\bibfnamefont{J.}~\bibnamefont{Chu}},
\bibinfo{author}{\bibfnamefont{S.~S.} \bibnamefont{Jaswal}},
\bibnamefont{and} \bibinfo{author}{\bibfnamefont{E.~Y.}
	\bibnamefont{Tsymbal}}, \bibinfo{journal}{Physical Review Letters}
\textbf{\bibinfo{volume}{101}}, \bibinfo{pages}{137201}
(\bibinfo{year}{2008}), ISSN \bibinfo{issn}{0031-9007, 1079-7114},
\urlprefix\url{https://link.aps.org/doi/10.1103/PhysRevLett.101.137201}.

\bibitem[{\citenamefont{Bernand-Mantel
		et~al.}(2013)\citenamefont{Bernand-Mantel, Herrera-Diez, Ranno, Pizzini,
		Vogel, Givord, Auffret, Boulle, Miron, and
		Gaudin}}]{bernand-mantel_electric-field_2013}
\bibinfo{author}{\bibfnamefont{A.}~\bibnamefont{Bernand-Mantel}},
\bibinfo{author}{\bibfnamefont{L.}~\bibnamefont{Herrera-Diez}},
\bibinfo{author}{\bibfnamefont{L.}~\bibnamefont{Ranno}},
\bibinfo{author}{\bibfnamefont{S.}~\bibnamefont{Pizzini}},
\bibinfo{author}{\bibfnamefont{J.}~\bibnamefont{Vogel}},
\bibinfo{author}{\bibfnamefont{D.}~\bibnamefont{Givord}},
\bibinfo{author}{\bibfnamefont{S.}~\bibnamefont{Auffret}},
\bibinfo{author}{\bibfnamefont{O.}~\bibnamefont{Boulle}},
\bibinfo{author}{\bibfnamefont{I.~M.} \bibnamefont{Miron}}, \bibnamefont{and}
\bibinfo{author}{\bibfnamefont{G.}~\bibnamefont{Gaudin}},
\bibinfo{journal}{Applied Physics Letters} \textbf{\bibinfo{volume}{102}},
\bibinfo{pages}{122406} (\bibinfo{year}{2013}), ISSN
\bibinfo{issn}{0003-6951, 1077-3118},
\urlprefix\url{http://aip.scitation.org/doi/10.1063/1.4798506}.

\bibitem[{\citenamefont{Liu et~al.}(2017)\citenamefont{Liu, Ono, Agnus, Adam,
		Jaiswal, Langer, Ocker, Ravelosona, and Herrera~Diez}}]{liu_electric_2017}
\bibinfo{author}{\bibfnamefont{Y.~T.} \bibnamefont{Liu}},
\bibinfo{author}{\bibfnamefont{S.}~\bibnamefont{Ono}},
\bibinfo{author}{\bibfnamefont{G.}~\bibnamefont{Agnus}},
\bibinfo{author}{\bibfnamefont{J.-P.} \bibnamefont{Adam}},
\bibinfo{author}{\bibfnamefont{S.}~\bibnamefont{Jaiswal}},
\bibinfo{author}{\bibfnamefont{J.}~\bibnamefont{Langer}},
\bibinfo{author}{\bibfnamefont{B.}~\bibnamefont{Ocker}},
\bibinfo{author}{\bibfnamefont{D.}~\bibnamefont{Ravelosona}},
\bibnamefont{and}
\bibinfo{author}{\bibfnamefont{L.}~\bibnamefont{Herrera~Diez}},
\bibinfo{journal}{Journal of Applied Physics} \textbf{\bibinfo{volume}{122}},
\bibinfo{pages}{133907} (\bibinfo{year}{2017}), ISSN
\bibinfo{issn}{0021-8979, 1089-7550},
\urlprefix\url{http://aip.scitation.org/doi/10.1063/1.4997834}.

\bibitem[{\citenamefont{Wang et~al.}(2012)\citenamefont{Wang, Li, Hageman, and
		Chien}}]{wang_electric-field-assisted_2012}
\bibinfo{author}{\bibfnamefont{W.-G.} \bibnamefont{Wang}},
\bibinfo{author}{\bibfnamefont{M.}~\bibnamefont{Li}},
\bibinfo{author}{\bibfnamefont{S.}~\bibnamefont{Hageman}}, \bibnamefont{and}
\bibinfo{author}{\bibfnamefont{C.~L.} \bibnamefont{Chien}},
\bibinfo{journal}{Nature Materials} \textbf{\bibinfo{volume}{11}},
\bibinfo{pages}{64} (\bibinfo{year}{2012}), ISSN \bibinfo{issn}{1476-1122,
	1476-4660}, \urlprefix\url{http://www.nature.com/articles/nmat3171}.

\bibitem[{\citenamefont{Hsu et~al.}(2017)\citenamefont{Hsu, Kubetzka, Finco,
		Romming, von Bergmann, and Wiesendanger}}]{hsu_electric-field-driven_2017}
\bibinfo{author}{\bibfnamefont{P.-J.} \bibnamefont{Hsu}},
\bibinfo{author}{\bibfnamefont{A.}~\bibnamefont{Kubetzka}},
\bibinfo{author}{\bibfnamefont{A.}~\bibnamefont{Finco}},
\bibinfo{author}{\bibfnamefont{N.}~\bibnamefont{Romming}},
\bibinfo{author}{\bibfnamefont{K.}~\bibnamefont{von Bergmann}},
\bibnamefont{and}
\bibinfo{author}{\bibfnamefont{R.}~\bibnamefont{Wiesendanger}},
\bibinfo{journal}{Nature Nanotechnology} \textbf{\bibinfo{volume}{12}},
\bibinfo{pages}{123} (\bibinfo{year}{2017}), ISSN \bibinfo{issn}{1748-3387,
	1748-3395}, \urlprefix\url{http://www.nature.com/articles/nnano.2016.234}.

\bibitem[{\citenamefont{Schott et~al.}(2017)\citenamefont{Schott,
		Bernand-Mantel, Ranno, Pizzini, Vogel, Béa, Baraduc, Auffret, Gaudin, and
		Givord}}]{schott_skyrmion_2017}
\bibinfo{author}{\bibfnamefont{M.}~\bibnamefont{Schott}},
\bibinfo{author}{\bibfnamefont{A.}~\bibnamefont{Bernand-Mantel}},
\bibinfo{author}{\bibfnamefont{L.}~\bibnamefont{Ranno}},
\bibinfo{author}{\bibfnamefont{S.}~\bibnamefont{Pizzini}},
\bibinfo{author}{\bibfnamefont{J.}~\bibnamefont{Vogel}},
\bibinfo{author}{\bibfnamefont{H.}~\bibnamefont{Béa}},
\bibinfo{author}{\bibfnamefont{C.}~\bibnamefont{Baraduc}},
\bibinfo{author}{\bibfnamefont{S.}~\bibnamefont{Auffret}},
\bibinfo{author}{\bibfnamefont{G.}~\bibnamefont{Gaudin}}, \bibnamefont{and}
\bibinfo{author}{\bibfnamefont{D.}~\bibnamefont{Givord}},
\bibinfo{journal}{Nano Letters} \textbf{\bibinfo{volume}{17}},
\bibinfo{pages}{3006} (\bibinfo{year}{2017}), ISSN \bibinfo{issn}{1530-6984,
	1530-6992},
\urlprefix\url{https://pubs.acs.org/doi/10.1021/acs.nanolett.7b00328}.

\bibitem[{\citenamefont{Srivastava et~al.}(2018)\citenamefont{Srivastava,
		Schott, Juge, Křižáková, Belmeguenai, Roussigné, Bernand-Mantel, Ranno,
		Pizzini, Chérif et~al.}}]{srivastava_large-voltage_2018}
\bibinfo{author}{\bibfnamefont{T.}~\bibnamefont{Srivastava}},
\bibinfo{author}{\bibfnamefont{M.}~\bibnamefont{Schott}},
\bibinfo{author}{\bibfnamefont{R.}~\bibnamefont{Juge}},
\bibinfo{author}{\bibfnamefont{V.}~\bibnamefont{Křižáková}},
\bibinfo{author}{\bibfnamefont{M.}~\bibnamefont{Belmeguenai}},
\bibinfo{author}{\bibfnamefont{Y.}~\bibnamefont{Roussigné}},
\bibinfo{author}{\bibfnamefont{A.}~\bibnamefont{Bernand-Mantel}},
\bibinfo{author}{\bibfnamefont{L.}~\bibnamefont{Ranno}},
\bibinfo{author}{\bibfnamefont{S.}~\bibnamefont{Pizzini}},
\bibinfo{author}{\bibfnamefont{S.-M.} \bibnamefont{Chérif}},
\bibnamefont{et~al.}, \bibinfo{journal}{Nano Letters}
\textbf{\bibinfo{volume}{18}}, \bibinfo{pages}{4871} (\bibinfo{year}{2018}),
ISSN \bibinfo{issn}{1530-6984, 1530-6992},
\urlprefix\url{https://pubs.acs.org/doi/10.1021/acs.nanolett.8b01502}.

\bibitem[{\citenamefont{Koyama et~al.}(2018)\citenamefont{Koyama, Nakatani,
		Ieda, and Chiba}}]{koyama_electric_2018}
\bibinfo{author}{\bibfnamefont{T.}~\bibnamefont{Koyama}},
\bibinfo{author}{\bibfnamefont{Y.}~\bibnamefont{Nakatani}},
\bibinfo{author}{\bibfnamefont{J.}~\bibnamefont{Ieda}}, \bibnamefont{and}
\bibinfo{author}{\bibfnamefont{D.}~\bibnamefont{Chiba}},
\bibinfo{journal}{Science Advances} \textbf{\bibinfo{volume}{4}},
\bibinfo{pages}{eaav0265} (\bibinfo{year}{2018}), ISSN
\bibinfo{issn}{2375-2548},
\urlprefix\url{https://advances.sciencemag.org/lookup/doi/10.1126/sciadv.aav0265}.

\bibitem[{\citenamefont{Zhang}(1999)}]{zhang_spin-dependent_1999}
\bibinfo{author}{\bibfnamefont{S.}~\bibnamefont{Zhang}},
\bibinfo{journal}{Physical Review Letters} \textbf{\bibinfo{volume}{83}},
\bibinfo{pages}{640} (\bibinfo{year}{1999}), ISSN \bibinfo{issn}{0031-9007,
	1079-7114},
\urlprefix\url{https://link.aps.org/doi/10.1103/PhysRevLett.83.640}.

\bibitem[{\citenamefont{Bauer et~al.}(2013)\citenamefont{Bauer, Emori, and
		Beach}}]{bauer_voltage-controlled_2013}
\bibinfo{author}{\bibfnamefont{U.}~\bibnamefont{Bauer}},
\bibinfo{author}{\bibfnamefont{S.}~\bibnamefont{Emori}}, \bibnamefont{and}
\bibinfo{author}{\bibfnamefont{G.~S.~D.} \bibnamefont{Beach}},
\bibinfo{journal}{Nature Nanotechnology} \textbf{\bibinfo{volume}{8}},
\bibinfo{pages}{411} (\bibinfo{year}{2013}), ISSN \bibinfo{issn}{1748-3387,
	1748-3395}, \urlprefix\url{http://www.nature.com/articles/nnano.2013.96}.

\bibitem[{\citenamefont{Bauer et~al.}(2015)\citenamefont{Bauer, Yao, Tan,
		Agrawal, Emori, Tuller, van Dijken, and Beach}}]{bauer_magneto-ionic_2015}
\bibinfo{author}{\bibfnamefont{U.}~\bibnamefont{Bauer}},
\bibinfo{author}{\bibfnamefont{L.}~\bibnamefont{Yao}},
\bibinfo{author}{\bibfnamefont{A.~J.} \bibnamefont{Tan}},
\bibinfo{author}{\bibfnamefont{P.}~\bibnamefont{Agrawal}},
\bibinfo{author}{\bibfnamefont{S.}~\bibnamefont{Emori}},
\bibinfo{author}{\bibfnamefont{H.~L.} \bibnamefont{Tuller}},
\bibinfo{author}{\bibfnamefont{S.}~\bibnamefont{van Dijken}},
\bibnamefont{and} \bibinfo{author}{\bibfnamefont{G.~S.~D.}
	\bibnamefont{Beach}}, \bibinfo{journal}{Nature Materials}
\textbf{\bibinfo{volume}{14}}, \bibinfo{pages}{174} (\bibinfo{year}{2015}),
ISSN \bibinfo{issn}{1476-1122, 1476-4660},
\urlprefix\url{http://www.nature.com/articles/nmat4134}.

\bibitem[{\citenamefont{Mishra et~al.}(2019)\citenamefont{Mishra, Mahfouzi,
		Kumar, Cai, Chen, Qiu, Kioussis, and Yang}}]{mishra_electric-field_2019}
\bibinfo{author}{\bibfnamefont{R.}~\bibnamefont{Mishra}},
\bibinfo{author}{\bibfnamefont{F.}~\bibnamefont{Mahfouzi}},
\bibinfo{author}{\bibfnamefont{D.}~\bibnamefont{Kumar}},
\bibinfo{author}{\bibfnamefont{K.}~\bibnamefont{Cai}},
\bibinfo{author}{\bibfnamefont{M.}~\bibnamefont{Chen}},
\bibinfo{author}{\bibfnamefont{X.}~\bibnamefont{Qiu}},
\bibinfo{author}{\bibfnamefont{N.}~\bibnamefont{Kioussis}}, \bibnamefont{and}
\bibinfo{author}{\bibfnamefont{H.}~\bibnamefont{Yang}},
\bibinfo{journal}{Nature Communications} \textbf{\bibinfo{volume}{10}},
\bibinfo{pages}{248} (\bibinfo{year}{2019}), ISSN \bibinfo{issn}{2041-1723},
\urlprefix\url{http://www.nature.com/articles/s41467-018-08274-8}.

\bibitem[{\citenamefont{Herrera~Diez et~al.}(2019)\citenamefont{Herrera~Diez,
		Liu, Gilbert, Belmeguenai, Vogel, Pizzini, Martinez, Lamperti, Mohammedi,
		Laborieux et~al.}}]{herrera_diez_nonvolatile_2019}
\bibinfo{author}{\bibfnamefont{L.}~\bibnamefont{Herrera~Diez}},
\bibinfo{author}{\bibfnamefont{Y.}~\bibnamefont{Liu}},
\bibinfo{author}{\bibfnamefont{D.}~\bibnamefont{Gilbert}},
\bibinfo{author}{\bibfnamefont{M.}~\bibnamefont{Belmeguenai}},
\bibinfo{author}{\bibfnamefont{J.}~\bibnamefont{Vogel}},
\bibinfo{author}{\bibfnamefont{S.}~\bibnamefont{Pizzini}},
\bibinfo{author}{\bibfnamefont{E.}~\bibnamefont{Martinez}},
\bibinfo{author}{\bibfnamefont{A.}~\bibnamefont{Lamperti}},
\bibinfo{author}{\bibfnamefont{J.}~\bibnamefont{Mohammedi}},
\bibinfo{author}{\bibfnamefont{A.}~\bibnamefont{Laborieux}},
\bibnamefont{et~al.}, \bibinfo{journal}{Physical Review Applied}
\textbf{\bibinfo{volume}{12}}, \bibinfo{pages}{034005}
(\bibinfo{year}{2019}), ISSN \bibinfo{issn}{2331-7019},
\urlprefix\url{https://link.aps.org/doi/10.1103/PhysRevApplied.12.034005}.

\bibitem[{\citenamefont{Chen et~al.}(2020)\citenamefont{Chen, Mascaraque, Jia,
		Zimmermann, Robertson, Conte, Hoffmann, Barrio, Ding, Wiesendanger
		et~al.}}]{chen_large_2020}
\bibinfo{author}{\bibfnamefont{G.}~\bibnamefont{Chen}},
\bibinfo{author}{\bibfnamefont{A.}~\bibnamefont{Mascaraque}},
\bibinfo{author}{\bibfnamefont{H.}~\bibnamefont{Jia}},
\bibinfo{author}{\bibfnamefont{B.}~\bibnamefont{Zimmermann}},
\bibinfo{author}{\bibfnamefont{M.}~\bibnamefont{Robertson}},
\bibinfo{author}{\bibfnamefont{R.~L.} \bibnamefont{Conte}},
\bibinfo{author}{\bibfnamefont{M.}~\bibnamefont{Hoffmann}},
\bibinfo{author}{\bibfnamefont{M.~A.~G.} \bibnamefont{Barrio}},
\bibinfo{author}{\bibfnamefont{H.}~\bibnamefont{Ding}},
\bibinfo{author}{\bibfnamefont{R.}~\bibnamefont{Wiesendanger}},
\bibnamefont{et~al.}, \bibinfo{journal}{Science Advances}
\textbf{\bibinfo{volume}{6}}, \bibinfo{pages}{eaba4924}
(\bibinfo{year}{2020}), ISSN \bibinfo{issn}{2375-2548},
\bibinfo{note}{publisher: American Association for the Advancement of Science
	Section: Research Article},
\urlprefix\url{https://advances.sciencemag.org/content/6/33/eaba4924}.

\bibitem[{\citenamefont{Gilbert et~al.}(2016)\citenamefont{Gilbert, Grutter,
		Arenholz, Liu, Kirby, Borchers, and Maranville}}]{gilbert_structural_2016}
\bibinfo{author}{\bibfnamefont{D.~A.} \bibnamefont{Gilbert}},
\bibinfo{author}{\bibfnamefont{A.~J.} \bibnamefont{Grutter}},
\bibinfo{author}{\bibfnamefont{E.}~\bibnamefont{Arenholz}},
\bibinfo{author}{\bibfnamefont{K.}~\bibnamefont{Liu}},
\bibinfo{author}{\bibfnamefont{B.~J.} \bibnamefont{Kirby}},
\bibinfo{author}{\bibfnamefont{J.~A.} \bibnamefont{Borchers}},
\bibnamefont{and} \bibinfo{author}{\bibfnamefont{B.~B.}
	\bibnamefont{Maranville}}, \bibinfo{journal}{Nature Communications}
\textbf{\bibinfo{volume}{7}}, \bibinfo{pages}{12264} (\bibinfo{year}{2016}),
ISSN \bibinfo{issn}{2041-1723},
\urlprefix\url{http://www.nature.com/articles/ncomms12264}.

\bibitem[{\citenamefont{Tan et~al.}(2019)\citenamefont{Tan, Huang, Avci,
		Büttner, Mann, Hu, Mazzoli, Wilkins, Tuller, and
		Beach}}]{tan_magneto-ionic_2019}
\bibinfo{author}{\bibfnamefont{A.~J.} \bibnamefont{Tan}},
\bibinfo{author}{\bibfnamefont{M.}~\bibnamefont{Huang}},
\bibinfo{author}{\bibfnamefont{C.~O.} \bibnamefont{Avci}},
\bibinfo{author}{\bibfnamefont{F.}~\bibnamefont{Büttner}},
\bibinfo{author}{\bibfnamefont{M.}~\bibnamefont{Mann}},
\bibinfo{author}{\bibfnamefont{W.}~\bibnamefont{Hu}},
\bibinfo{author}{\bibfnamefont{C.}~\bibnamefont{Mazzoli}},
\bibinfo{author}{\bibfnamefont{S.}~\bibnamefont{Wilkins}},
\bibinfo{author}{\bibfnamefont{H.~L.} \bibnamefont{Tuller}},
\bibnamefont{and} \bibinfo{author}{\bibfnamefont{G.~S.~D.}
	\bibnamefont{Beach}}, \bibinfo{journal}{Nature Materials}
\textbf{\bibinfo{volume}{18}}, \bibinfo{pages}{35} (\bibinfo{year}{2019}),
ISSN \bibinfo{issn}{1476-1122, 1476-4660},
\urlprefix\url{http://www.nature.com/articles/s41563-018-0211-5}.

\bibitem[{\citenamefont{Lee et~al.}(2020)\citenamefont{Lee, Jo, Tan, Huang,
		Choi, Park, Ji, Son, Chang, Beach et~al.}}]{lee_fast_2020}
\bibinfo{author}{\bibfnamefont{K.-Y.} \bibnamefont{Lee}},
\bibinfo{author}{\bibfnamefont{S.}~\bibnamefont{Jo}},
\bibinfo{author}{\bibfnamefont{A.~J.} \bibnamefont{Tan}},
\bibinfo{author}{\bibfnamefont{M.}~\bibnamefont{Huang}},
\bibinfo{author}{\bibfnamefont{D.}~\bibnamefont{Choi}},
\bibinfo{author}{\bibfnamefont{J.~H.} \bibnamefont{Park}},
\bibinfo{author}{\bibfnamefont{H.-I.} \bibnamefont{Ji}},
\bibinfo{author}{\bibfnamefont{J.-W.} \bibnamefont{Son}},
\bibinfo{author}{\bibfnamefont{J.}~\bibnamefont{Chang}},
\bibinfo{author}{\bibfnamefont{G.~S.~D.} \bibnamefont{Beach}},
\bibnamefont{et~al.}, \bibinfo{journal}{Nano Letters}
\textbf{\bibinfo{volume}{20}}, \bibinfo{pages}{3435} (\bibinfo{year}{2020}),
ISSN \bibinfo{issn}{1530-6984, 1530-6992},
\urlprefix\url{https://pubs.acs.org/doi/10.1021/acs.nanolett.0c00340}.

\bibitem[{\citenamefont{Fassatoui}(2020)}]{fassatoui_physical_2020}
\bibinfo{author}{\bibfnamefont{A.}~\bibnamefont{Fassatoui}},
\bibinfo{journal}{Physical Review Applied}  (\bibinfo{year}{2020}),
\urlprefix\url{https://journals.aps.org/prapplied/accepted/e707fA70H6410103f2932974e2842efaacadb08cd}.

\bibitem[{\citenamefont{Leighton}(2019)}]{leighton_ionic-control_2019}
\bibinfo{author}{\bibfnamefont{C.}~\bibnamefont{Leighton}},
\bibinfo{journal}{Nature Materials} \textbf{\bibinfo{volume}{18}},
\bibinfo{pages}{13} (\bibinfo{year}{2019}), ISSN \bibinfo{issn}{1476-1122,
	1476-4660}, \urlprefix\url{https://www.nature.com/articles/ncomms3676}.

\bibitem[{\citenamefont{Jeong et~al.}(2013)\citenamefont{Jeong, Aetukuri, Graf,
		Schladt, Samant, and Parkin}}]{jeong_oxygen-vacancy_2013}
\bibinfo{author}{\bibfnamefont{J.}~\bibnamefont{Jeong}},
\bibinfo{author}{\bibfnamefont{N.}~\bibnamefont{Aetukuri}},
\bibinfo{author}{\bibfnamefont{T.}~\bibnamefont{Graf}},
\bibinfo{author}{\bibfnamefont{T.~D.} \bibnamefont{Schladt}},
\bibinfo{author}{\bibfnamefont{M.~G.} \bibnamefont{Samant}},
\bibnamefont{and} \bibinfo{author}{\bibfnamefont{S.~S.~P.}
	\bibnamefont{Parkin}}, \bibinfo{journal}{Science}
\textbf{\bibinfo{volume}{339}}, \bibinfo{pages}{1402} (\bibinfo{year}{2013}),
ISSN \bibinfo{issn}{1095-9203},
\urlprefix\url{https://science.sciencemag.org/content/339/6126/1402}.

\bibitem[{\citenamefont{Shi et~al.}(2013)\citenamefont{Shi, Ha, Zhou, Schoofs,
		and Ramanathan}}]{shi_resistance_2013}
\bibinfo{author}{\bibfnamefont{J.}~\bibnamefont{Shi}},
\bibinfo{author}{\bibfnamefont{S.~D.} \bibnamefont{Ha}},
\bibinfo{author}{\bibfnamefont{Y.}~\bibnamefont{Zhou}},
\bibinfo{author}{\bibfnamefont{F.}~\bibnamefont{Schoofs}}, \bibnamefont{and}
\bibinfo{author}{\bibfnamefont{S.}~\bibnamefont{Ramanathan}},
\bibinfo{journal}{Nature Communications} \textbf{\bibinfo{volume}{4}},
\bibinfo{pages}{2676} (\bibinfo{year}{2013}), ISSN \bibinfo{issn}{2041-1723},
\urlprefix\url{https://www.nature.com/articles/ncomms3676}.

\bibitem[{\citenamefont{Ono et~al.}(2008)\citenamefont{Ono, Seki, Hirahara,
		Tominari, and Takeya}}]{ono_ionic-liquid_2008}
\bibinfo{author}{\bibfnamefont{S.}~\bibnamefont{Ono}},
\bibinfo{author}{\bibfnamefont{S.}~\bibnamefont{Seki}},
\bibinfo{author}{\bibfnamefont{R.}~\bibnamefont{Hirahara}},
\bibinfo{author}{\bibfnamefont{Y.}~\bibnamefont{Tominari}}, \bibnamefont{and}
\bibinfo{author}{\bibfnamefont{J.}~\bibnamefont{Takeya}},
\bibinfo{journal}{Applied Physics Letters} \textbf{\bibinfo{volume}{92}},
\bibinfo{pages}{103313} (\bibinfo{year}{2008}), ISSN
\bibinfo{issn}{0003-6951}, \bibinfo{note}{publisher: American Institute of
	Physics}, \urlprefix\url{https://aip.scitation.org/doi/10.1063/1.2898203}.

\bibitem[{\citenamefont{Manchon et~al.}(2008)\citenamefont{Manchon, Ducruet,
		Lombard, Auffret, Rodmacq, Dieny, Pizzini, Vogel, and
		Uhlí}}]{manchon_analysis_2008}
\bibinfo{author}{\bibfnamefont{A.}~\bibnamefont{Manchon}},
\bibinfo{author}{\bibfnamefont{C.}~\bibnamefont{Ducruet}},
\bibinfo{author}{\bibfnamefont{L.}~\bibnamefont{Lombard}},
\bibinfo{author}{\bibfnamefont{S.}~\bibnamefont{Auffret}},
\bibinfo{author}{\bibfnamefont{B.}~\bibnamefont{Rodmacq}},
\bibinfo{author}{\bibfnamefont{B.}~\bibnamefont{Dieny}},
\bibinfo{author}{\bibfnamefont{S.}~\bibnamefont{Pizzini}},
\bibinfo{author}{\bibfnamefont{J.}~\bibnamefont{Vogel}}, \bibnamefont{and}
\bibinfo{author}{\bibfnamefont{V.}~\bibnamefont{Uhlí}}, \bibinfo{journal}{J.
	Appl. Phys.} p.~\bibinfo{pages}{8} (\bibinfo{year}{2008}).

\bibitem[{\citenamefont{Nagata et~al.}(2010)\citenamefont{Nagata, Haemori,
		Yamashita, Yoshikawa, Iwashita, Kobayashi, and Chikyow}}]{nagata_oxygen_2010}
\bibinfo{author}{\bibfnamefont{T.}~\bibnamefont{Nagata}},
\bibinfo{author}{\bibfnamefont{M.}~\bibnamefont{Haemori}},
\bibinfo{author}{\bibfnamefont{Y.}~\bibnamefont{Yamashita}},
\bibinfo{author}{\bibfnamefont{H.}~\bibnamefont{Yoshikawa}},
\bibinfo{author}{\bibfnamefont{Y.}~\bibnamefont{Iwashita}},
\bibinfo{author}{\bibfnamefont{K.}~\bibnamefont{Kobayashi}},
\bibnamefont{and} \bibinfo{author}{\bibfnamefont{T.}~\bibnamefont{Chikyow}},
\bibinfo{journal}{Applied Physics Letters} \textbf{\bibinfo{volume}{97}},
\bibinfo{pages}{082902} (\bibinfo{year}{2010}), ISSN
\bibinfo{issn}{0003-6951}, \bibinfo{note}{publisher: American Institute of
	PhysicsAIP},
\urlprefix\url{https://aip-scitation-org.proxy.scd.u-psud.fr/doi/abs/10.1063/1.3483756}.

\bibitem[{\citenamefont{Yuasa et~al.}(2005)\citenamefont{Yuasa, Suzuki,
		Katayama, and Ando}}]{yuasa_characterization_2005}
\bibinfo{author}{\bibfnamefont{S.}~\bibnamefont{Yuasa}},
\bibinfo{author}{\bibfnamefont{Y.}~\bibnamefont{Suzuki}},
\bibinfo{author}{\bibfnamefont{T.}~\bibnamefont{Katayama}}, \bibnamefont{and}
\bibinfo{author}{\bibfnamefont{K.}~\bibnamefont{Ando}},
\bibinfo{journal}{Applied Physics Letters} \textbf{\bibinfo{volume}{87}},
\bibinfo{pages}{242503} (\bibinfo{year}{2005}), ISSN
\bibinfo{issn}{0003-6951}, \bibinfo{note}{publisher: American Institute of
	Physics},
\urlprefix\url{https://aip.scitation.org/doi/citedby/10.1063/1.2140612}.

\bibitem[{\citenamefont{Wang et~al.}(2009)\citenamefont{Wang, Jordan-sweet,
		Miao, Ni, Rumaiz, Shah, Fan, Parsons, Stearrett, Nowak
		et~al.}}]{wang_-situ_2009}
\bibinfo{author}{\bibfnamefont{W.~G.} \bibnamefont{Wang}},
\bibinfo{author}{\bibfnamefont{J.}~\bibnamefont{Jordan-sweet}},
\bibinfo{author}{\bibfnamefont{G.~X.} \bibnamefont{Miao}},
\bibinfo{author}{\bibfnamefont{C.}~\bibnamefont{Ni}},
\bibinfo{author}{\bibfnamefont{A.~K.} \bibnamefont{Rumaiz}},
\bibinfo{author}{\bibfnamefont{L.~R.} \bibnamefont{Shah}},
\bibinfo{author}{\bibfnamefont{X.}~\bibnamefont{Fan}},
\bibinfo{author}{\bibfnamefont{P.}~\bibnamefont{Parsons}},
\bibinfo{author}{\bibfnamefont{R.}~\bibnamefont{Stearrett}},
\bibinfo{author}{\bibfnamefont{E.~R.} \bibnamefont{Nowak}},
\bibnamefont{et~al.}, \bibinfo{journal}{Applied Physics Letters}
\textbf{\bibinfo{volume}{95}}, \bibinfo{pages}{242501}
(\bibinfo{year}{2009}), ISSN \bibinfo{issn}{0003-6951},
\bibinfo{note}{publisher: American Institute of Physics},
\urlprefix\url{https://aip.scitation.org/doi/abs/10.1063/1.3273397}.

\bibitem[{\citenamefont{Sinha et~al.}(2015)\citenamefont{Sinha, Gruber,
		Kodzuka, Ohkubo, Mitani, Hono, and Hayashi}}]{sinha_borondiffusion_2015}
\bibinfo{author}{\bibfnamefont{J.}~\bibnamefont{Sinha}},
\bibinfo{author}{\bibfnamefont{M.}~\bibnamefont{Gruber}},
\bibinfo{author}{\bibfnamefont{M.}~\bibnamefont{Kodzuka}},
\bibinfo{author}{\bibfnamefont{T.}~\bibnamefont{Ohkubo}},
\bibinfo{author}{\bibfnamefont{S.}~\bibnamefont{Mitani}},
\bibinfo{author}{\bibfnamefont{K.}~\bibnamefont{Hono}}, \bibnamefont{and}
\bibinfo{author}{\bibfnamefont{M.}~\bibnamefont{Hayashi}},
\bibinfo{journal}{J. Appl. Phys.} \textbf{\bibinfo{volume}{117}},
\bibinfo{pages}{043913} (\bibinfo{year}{2015}).

\bibitem[{\citenamefont{Vermeulen et~al.}(2017)\citenamefont{Vermeulen, Wu,
		Swerts, Couet, Radu, Groeseneken, Detavernier, Jochum, Bael, Temst
		et~al.}}]{bart_hfo2_2017}
\bibinfo{author}{\bibfnamefont{B.~F.} \bibnamefont{Vermeulen}},
\bibinfo{author}{\bibfnamefont{J.}~\bibnamefont{Wu}},
\bibinfo{author}{\bibfnamefont{J.}~\bibnamefont{Swerts}},
\bibinfo{author}{\bibfnamefont{S.}~\bibnamefont{Couet}},
\bibinfo{author}{\bibfnamefont{I.~P.} \bibnamefont{Radu}},
\bibinfo{author}{\bibfnamefont{G.}~\bibnamefont{Groeseneken}},
\bibinfo{author}{\bibfnamefont{C.}~\bibnamefont{Detavernier}},
\bibinfo{author}{\bibfnamefont{J.~K.} \bibnamefont{Jochum}},
\bibinfo{author}{\bibfnamefont{M.~V.} \bibnamefont{Bael}},
\bibinfo{author}{\bibfnamefont{K.}~\bibnamefont{Temst}},
\bibnamefont{et~al.}, \bibinfo{journal}{AIP Advances}
\textbf{\bibinfo{volume}{7}}, \bibinfo{pages}{055933} (\bibinfo{year}{2017}).

\bibitem[{\citenamefont{Seidemann}(2017)}]{seidemann_iontronics_2017}
\bibinfo{author}{\bibfnamefont{J.}~\bibnamefont{Seidemann}}, Ph.D. thesis,
\bibinfo{school}{Université Grenoble Alpes} (\bibinfo{year}{2017}),
\urlprefix\url{https://tel.archives-ouvertes.fr/tel-01759252}.

\bibitem[{\citenamefont{Belmeguenai et~al.}(2015)\citenamefont{Belmeguenai,
		Adam, Roussigné, Eimer, Devolder, Kim, Cherif, Stashkevich, and
		Thiaville}}]{belmeguenai_interfacial_2015}
\bibinfo{author}{\bibfnamefont{M.}~\bibnamefont{Belmeguenai}},
\bibinfo{author}{\bibfnamefont{J.-P.} \bibnamefont{Adam}},
\bibinfo{author}{\bibfnamefont{Y.}~\bibnamefont{Roussigné}},
\bibinfo{author}{\bibfnamefont{S.}~\bibnamefont{Eimer}},
\bibinfo{author}{\bibfnamefont{T.}~\bibnamefont{Devolder}},
\bibinfo{author}{\bibfnamefont{J.-V.} \bibnamefont{Kim}},
\bibinfo{author}{\bibfnamefont{S.~M.} \bibnamefont{Cherif}},
\bibinfo{author}{\bibfnamefont{A.}~\bibnamefont{Stashkevich}},
\bibnamefont{and}
\bibinfo{author}{\bibfnamefont{A.}~\bibnamefont{Thiaville}},
\bibinfo{journal}{Physical Review B} \textbf{\bibinfo{volume}{91}},
\bibinfo{pages}{180405} (\bibinfo{year}{2015}), ISSN
\bibinfo{issn}{1098-0121, 1550-235X},
\urlprefix\url{https://link.aps.org/doi/10.1103/PhysRevB.91.180405}.

\bibitem[{\citenamefont{Nakayama et~al.}(2012)\citenamefont{Nakayama, Ando,
		Harii, Yoshino, Takahashi, Kajiwara, Uchida, Fujikawa, and
		Saitoh}}]{nakayama_geometry_2012}
\bibinfo{author}{\bibfnamefont{H.}~\bibnamefont{Nakayama}},
\bibinfo{author}{\bibfnamefont{K.}~\bibnamefont{Ando}},
\bibinfo{author}{\bibfnamefont{K.}~\bibnamefont{Harii}},
\bibinfo{author}{\bibfnamefont{T.}~\bibnamefont{Yoshino}},
\bibinfo{author}{\bibfnamefont{R.}~\bibnamefont{Takahashi}},
\bibinfo{author}{\bibfnamefont{Y.}~\bibnamefont{Kajiwara}},
\bibinfo{author}{\bibfnamefont{K.}~\bibnamefont{Uchida}},
\bibinfo{author}{\bibfnamefont{Y.}~\bibnamefont{Fujikawa}}, \bibnamefont{and}
\bibinfo{author}{\bibfnamefont{E.}~\bibnamefont{Saitoh}},
\bibinfo{journal}{Physical Review B} \textbf{\bibinfo{volume}{85}},
\bibinfo{pages}{144408} (\bibinfo{year}{2012}), \bibinfo{note}{publisher:
	American Physical Society},
\urlprefix\url{https://link.aps.org/doi/10.1103/PhysRevB.85.144408}.

\bibitem[{\citenamefont{Belmeguenai et~al.}(2018)\citenamefont{Belmeguenai,
		Gabor, Zighem, Challab, Petrisor, Mos, and
		Tiusan}}]{belmeguenai_ferromagnetic-resonance-induced_2018}
\bibinfo{author}{\bibfnamefont{M.}~\bibnamefont{Belmeguenai}},
\bibinfo{author}{\bibfnamefont{M.~S.} \bibnamefont{Gabor}},
\bibinfo{author}{\bibfnamefont{F.}~\bibnamefont{Zighem}},
\bibinfo{author}{\bibfnamefont{N.}~\bibnamefont{Challab}},
\bibinfo{author}{\bibfnamefont{T.}~\bibnamefont{Petrisor}},
\bibinfo{author}{\bibfnamefont{R.~B.} \bibnamefont{Mos}}, \bibnamefont{and}
\bibinfo{author}{\bibfnamefont{C.}~\bibnamefont{Tiusan}},
\bibinfo{journal}{Journal of Physics D: Applied Physics}
\textbf{\bibinfo{volume}{51}}, \bibinfo{pages}{045002}
(\bibinfo{year}{2018}), ISSN \bibinfo{issn}{0022-3727, 1361-6463},
\urlprefix\url{https://iopscience.iop.org/article/10.1088/1361-6463/aa9f55}.

\bibitem[{\citenamefont{Belmeguenai et~al.}(2017)\citenamefont{Belmeguenai,
		Apalkov, Roussigné, Chérif, Stashkevich, Feng, and
		Tang}}]{belmeguenai_exchange_2017}
\bibinfo{author}{\bibfnamefont{M.}~\bibnamefont{Belmeguenai}},
\bibinfo{author}{\bibfnamefont{D.}~\bibnamefont{Apalkov}},
\bibinfo{author}{\bibfnamefont{Y.}~\bibnamefont{Roussigné}},
\bibinfo{author}{\bibfnamefont{M.}~\bibnamefont{Chérif}},
\bibinfo{author}{\bibfnamefont{A.}~\bibnamefont{Stashkevich}},
\bibinfo{author}{\bibfnamefont{G.}~\bibnamefont{Feng}}, \bibnamefont{and}
\bibinfo{author}{\bibfnamefont{X.}~\bibnamefont{Tang}},
\bibinfo{journal}{Journal of Physics D: Applied Physics}
\textbf{\bibinfo{volume}{50}}, \bibinfo{pages}{415003}
(\bibinfo{year}{2017}), ISSN \bibinfo{issn}{0022-3727},
\bibinfo{note}{publisher: IOP Publishing},
\urlprefix\url{https://doi.org/10.1088/1361-6463/aa81a5}.

\bibitem[{\citenamefont{Kuanr et~al.}(2005)\citenamefont{Kuanr, Camley, and
		Celinski}}]{kuanr_extrinsic_2005}
\bibinfo{author}{\bibfnamefont{B.~K.} \bibnamefont{Kuanr}},
\bibinfo{author}{\bibfnamefont{R.~E.} \bibnamefont{Camley}},
\bibnamefont{and} \bibinfo{author}{\bibfnamefont{Z.}~\bibnamefont{Celinski}},
\bibinfo{journal}{Journal of Magnetism and Magnetic Materials}
\textbf{\bibinfo{volume}{286}}, \bibinfo{pages}{276} (\bibinfo{year}{2005}),
ISSN \bibinfo{issn}{0304-8853},
\urlprefix\url{http://www.sciencedirect.com/science/article/pii/S0304885304009503}.

\bibitem[{\citenamefont{Capua et~al.}(2015)\citenamefont{Capua, Yang, Phung,
		and Parkin}}]{capua_determination_2015}
\bibinfo{author}{\bibfnamefont{A.}~\bibnamefont{Capua}},
\bibinfo{author}{\bibfnamefont{S.-h.} \bibnamefont{Yang}},
\bibinfo{author}{\bibfnamefont{T.}~\bibnamefont{Phung}}, \bibnamefont{and}
\bibinfo{author}{\bibfnamefont{S.~S.~P.} \bibnamefont{Parkin}},
\bibinfo{journal}{Physical Review B} \textbf{\bibinfo{volume}{92}},
\bibinfo{pages}{224402} (\bibinfo{year}{2015}), \bibinfo{note}{publisher:
	American Physical Society},
\urlprefix\url{https://link.aps.org/doi/10.1103/PhysRevB.92.224402}.

\bibitem[{\citenamefont{Diez et~al.}(2019)\citenamefont{Diez, Voto, Casiraghi,
		Belmeguenai, Roussigné, Durin, Lamperti, Mantovan, Sluka, Jeudy
		et~al.}}]{diez_enhancement_2019}
\bibinfo{author}{\bibfnamefont{L.~H.} \bibnamefont{Diez}},
\bibinfo{author}{\bibfnamefont{M.}~\bibnamefont{Voto}},
\bibinfo{author}{\bibfnamefont{A.}~\bibnamefont{Casiraghi}},
\bibinfo{author}{\bibfnamefont{M.}~\bibnamefont{Belmeguenai}},
\bibinfo{author}{\bibfnamefont{Y.}~\bibnamefont{Roussigné}},
\bibinfo{author}{\bibfnamefont{G.}~\bibnamefont{Durin}},
\bibinfo{author}{\bibfnamefont{A.}~\bibnamefont{Lamperti}},
\bibinfo{author}{\bibfnamefont{R.}~\bibnamefont{Mantovan}},
\bibinfo{author}{\bibfnamefont{V.}~\bibnamefont{Sluka}},
\bibinfo{author}{\bibfnamefont{V.}~\bibnamefont{Jeudy}},
\bibnamefont{et~al.}, \bibinfo{journal}{Physical Review B}
\textbf{\bibinfo{volume}{99}}, \bibinfo{pages}{054431}
(\bibinfo{year}{2019}), ISSN \bibinfo{issn}{2469-9950, 2469-9969},
\urlprefix\url{https://link.aps.org/doi/10.1103/PhysRevB.99.054431}.

\bibitem[{\citenamefont{Yang et~al.}(2017)\citenamefont{Yang, Zhang, Jiang,
		Dong, Wang, Liu, Zhao, Wang, Sun, and Yu}}]{yang_effect_2017}
\bibinfo{author}{\bibfnamefont{G.}~\bibnamefont{Yang}},
\bibinfo{author}{\bibfnamefont{J.-Y.} \bibnamefont{Zhang}},
\bibinfo{author}{\bibfnamefont{S.-L.} \bibnamefont{Jiang}},
\bibinfo{author}{\bibfnamefont{B.-W.} \bibnamefont{Dong}},
\bibinfo{author}{\bibfnamefont{S.-G.} \bibnamefont{Wang}},
\bibinfo{author}{\bibfnamefont{J.-L.} \bibnamefont{Liu}},
\bibinfo{author}{\bibfnamefont{Y.-C.} \bibnamefont{Zhao}},
\bibinfo{author}{\bibfnamefont{C.}~\bibnamefont{Wang}},
\bibinfo{author}{\bibfnamefont{Y.}~\bibnamefont{Sun}}, \bibnamefont{and}
\bibinfo{author}{\bibfnamefont{G.-H.} \bibnamefont{Yu}},
\bibinfo{journal}{Applied Surface Science} \textbf{\bibinfo{volume}{396}},
\bibinfo{pages}{705} (\bibinfo{year}{2017}), ISSN \bibinfo{issn}{0169-4332},
\urlprefix\url{http://www.sciencedirect.com/science/article/pii/S0169433216323674}.

\bibitem[{\citenamefont{Ikeda et~al.}(2010)\citenamefont{Ikeda, Miura,
		Yamamoto, Mizunuma, Gan, Endo, Kanai, Hayakawa, Matsukura, and
		Ohno}}]{ikeda_perpendicular-anisotropy_2010}
\bibinfo{author}{\bibfnamefont{S.}~\bibnamefont{Ikeda}},
\bibinfo{author}{\bibfnamefont{K.}~\bibnamefont{Miura}},
\bibinfo{author}{\bibfnamefont{H.}~\bibnamefont{Yamamoto}},
\bibinfo{author}{\bibfnamefont{K.}~\bibnamefont{Mizunuma}},
\bibinfo{author}{\bibfnamefont{H.~D.} \bibnamefont{Gan}},
\bibinfo{author}{\bibfnamefont{M.}~\bibnamefont{Endo}},
\bibinfo{author}{\bibfnamefont{S.}~\bibnamefont{Kanai}},
\bibinfo{author}{\bibfnamefont{J.}~\bibnamefont{Hayakawa}},
\bibinfo{author}{\bibfnamefont{F.}~\bibnamefont{Matsukura}},
\bibnamefont{and} \bibinfo{author}{\bibfnamefont{H.}~\bibnamefont{Ohno}},
\bibinfo{journal}{Nature Materials} \textbf{\bibinfo{volume}{9}},
\bibinfo{pages}{721} (\bibinfo{year}{2010}), ISSN \bibinfo{issn}{1476-1122,
	1476-4660}, \urlprefix\url{http://www.nature.com/articles/nmat2804}.

\bibitem[{\citenamefont{Benguettat-El~Mokhtari
		et~al.}(2020)\citenamefont{Benguettat-El~Mokhtari, Roussigné, Chérif,
		Stashkevich, Auffret, Baraduc, Gabor, Béa, and
		Belmeguenai}}]{benguettat-el_mokhtari_interface_2020}
\bibinfo{author}{\bibfnamefont{I.}~\bibnamefont{Benguettat-El~Mokhtari}},
\bibinfo{author}{\bibfnamefont{Y.}~\bibnamefont{Roussigné}},
\bibinfo{author}{\bibfnamefont{S.~M.} \bibnamefont{Chérif}},
\bibinfo{author}{\bibfnamefont{A.}~\bibnamefont{Stashkevich}},
\bibinfo{author}{\bibfnamefont{S.}~\bibnamefont{Auffret}},
\bibinfo{author}{\bibfnamefont{C.}~\bibnamefont{Baraduc}},
\bibinfo{author}{\bibfnamefont{M.}~\bibnamefont{Gabor}},
\bibinfo{author}{\bibfnamefont{H.}~\bibnamefont{Béa}}, \bibnamefont{and}
\bibinfo{author}{\bibfnamefont{M.}~\bibnamefont{Belmeguenai}},
\bibinfo{journal}{Physical Review Materials} \textbf{\bibinfo{volume}{4}},
\bibinfo{pages}{124408} (\bibinfo{year}{2020}), \bibinfo{note}{publisher:
	American Physical Society},
\urlprefix\url{https://link.aps.org/doi/10.1103/PhysRevMaterials.4.124408}.

\bibitem[{\citenamefont{Iihama et~al.}(2014)\citenamefont{Iihama, Mizukami,
		Naganuma, Oogane, Ando, and Miyazaki}}]{iihama_gilbert_2014}
\bibinfo{author}{\bibfnamefont{S.}~\bibnamefont{Iihama}},
\bibinfo{author}{\bibfnamefont{S.}~\bibnamefont{Mizukami}},
\bibinfo{author}{\bibfnamefont{H.}~\bibnamefont{Naganuma}},
\bibinfo{author}{\bibfnamefont{M.}~\bibnamefont{Oogane}},
\bibinfo{author}{\bibfnamefont{Y.}~\bibnamefont{Ando}}, \bibnamefont{and}
\bibinfo{author}{\bibfnamefont{T.}~\bibnamefont{Miyazaki}},
\bibinfo{journal}{Physical Review B} \textbf{\bibinfo{volume}{89}}
(\bibinfo{year}{2014}).

\bibitem[{\citenamefont{Lourembam et~al.}(2018)\citenamefont{Lourembam, Ghosh,
		Zeng, Wong, Yap, and Ter~Lim}}]{lourembam_thickness-dependent_2018}
\bibinfo{author}{\bibfnamefont{J.}~\bibnamefont{Lourembam}},
\bibinfo{author}{\bibfnamefont{A.}~\bibnamefont{Ghosh}},
\bibinfo{author}{\bibfnamefont{M.}~\bibnamefont{Zeng}},
\bibinfo{author}{\bibfnamefont{S.~K.} \bibnamefont{Wong}},
\bibinfo{author}{\bibfnamefont{Q.~J.} \bibnamefont{Yap}}, \bibnamefont{and}
\bibinfo{author}{\bibfnamefont{S.}~\bibnamefont{Ter~Lim}},
\bibinfo{journal}{Physical Review Applied} \textbf{\bibinfo{volume}{10}},
\bibinfo{pages}{044057} (\bibinfo{year}{2018}), ISSN
\bibinfo{issn}{2331-7019},
\urlprefix\url{https://link.aps.org/doi/10.1103/PhysRevApplied.10.044057}.

\bibitem[{\citenamefont{Bridge and Lambert}(1979)}]{bridge_oxygen_1979}
\bibinfo{author}{\bibfnamefont{M.~E.} \bibnamefont{Bridge}} \bibnamefont{and}
\bibinfo{author}{\bibfnamefont{R.~M.} \bibnamefont{Lambert}},
\bibinfo{journal}{Surface Science} \textbf{\bibinfo{volume}{82}},
\bibinfo{pages}{413} (\bibinfo{year}{1979}), ISSN \bibinfo{issn}{0039-6028},
\urlprefix\url{http://www.sciencedirect.com/science/article/pii/0039602879901997}.

\bibitem[{\citenamefont{Brundle}(1977)}]{brundle_oxygen_1977}
\bibinfo{author}{\bibfnamefont{C.}~\bibnamefont{Brundle}},
\bibinfo{journal}{Surface Science} \textbf{\bibinfo{volume}{66}},
\bibinfo{pages}{581} (\bibinfo{year}{1977}), ISSN \bibinfo{issn}{00396028},
\urlprefix\url{https://linkinghub.elsevier.com/retrieve/pii/0039602877900395}.

\bibitem[{\citenamefont{Simmons and Dwyer}(1975)}]{simmons_leed-aes_1975}
\bibinfo{author}{\bibfnamefont{G.~W.} \bibnamefont{Simmons}} \bibnamefont{and}
\bibinfo{author}{\bibfnamefont{D.~J.} \bibnamefont{Dwyer}},
\bibinfo{journal}{Surface Science} \textbf{\bibinfo{volume}{48}},
\bibinfo{pages}{373} (\bibinfo{year}{1975}), ISSN \bibinfo{issn}{00396028},
\urlprefix\url{https://linkinghub.elsevier.com/retrieve/pii/0039602875904136}.
	
\bibitem[{\citenamefont{Herrera~Diez et~al.}(2015)\citenamefont{Herrera~Diez,
		García-Sánchez, Adam, Devolder, Eimer, El~Hadri, Lamperti, Mantovan, Ocker,
		and Ravelosona}}]{herrera_diez_controlling_2015}
\bibinfo{author}{\bibfnamefont{L.}~\bibnamefont{Herrera~Diez}},
\bibinfo{author}{\bibfnamefont{F.}~\bibnamefont{García-Sánchez}},
\bibinfo{author}{\bibfnamefont{J.-P.} \bibnamefont{Adam}},
\bibinfo{author}{\bibfnamefont{T.}~\bibnamefont{Devolder}},
\bibinfo{author}{\bibfnamefont{S.}~\bibnamefont{Eimer}},
\bibinfo{author}{\bibfnamefont{M.~S.} \bibnamefont{El~Hadri}},
\bibinfo{author}{\bibfnamefont{A.}~\bibnamefont{Lamperti}},
\bibinfo{author}{\bibfnamefont{R.}~\bibnamefont{Mantovan}},
\bibinfo{author}{\bibfnamefont{B.}~\bibnamefont{Ocker}}, \bibnamefont{and}
\bibinfo{author}{\bibfnamefont{D.}~\bibnamefont{Ravelosona}},
\bibinfo{journal}{Applied Physics Letters} \textbf{\bibinfo{volume}{107}},
\bibinfo{pages}{032401} (\bibinfo{year}{2015}), ISSN
\bibinfo{issn}{0003-6951}, \bibinfo{note}{publisher: American Institute of
	Physics},
\urlprefix\url{http://aip.scitation.org/doi/full/10.1063/1.4927204}.

 

	
\end{thebibliography}

\end{document}